\PassOptionsToPackage{section,below}{placeins}
\documentclass[a4paper]{bytedance_seed}

\usepackage[utf8]{inputenc}
\usepackage{url}
\usepackage{booktabs}
\usepackage{amsfonts}
\usepackage{amsmath}
\usepackage{amssymb}
\usepackage{nicefrac}
\usepackage{graphicx}
\usepackage{xcolor}
\usepackage{multirow}
\usepackage{float}
\usepackage{tabularx}
\usepackage{subcaption}
\usepackage{oridynamic_style}

\ifpdftex
  \usepackage{tgpagella}
  \usepackage[T1]{fontenc}
\else
  \usepackage{unicode-math}
\fi

\hypersetup{citecolor=black}
\graphicspath{{figs/}}

\DeclareTextFontCommand{\textbf}{\bfseries}

\title{\raisebox{-0.22em}{\includegraphics[height=1.25em]{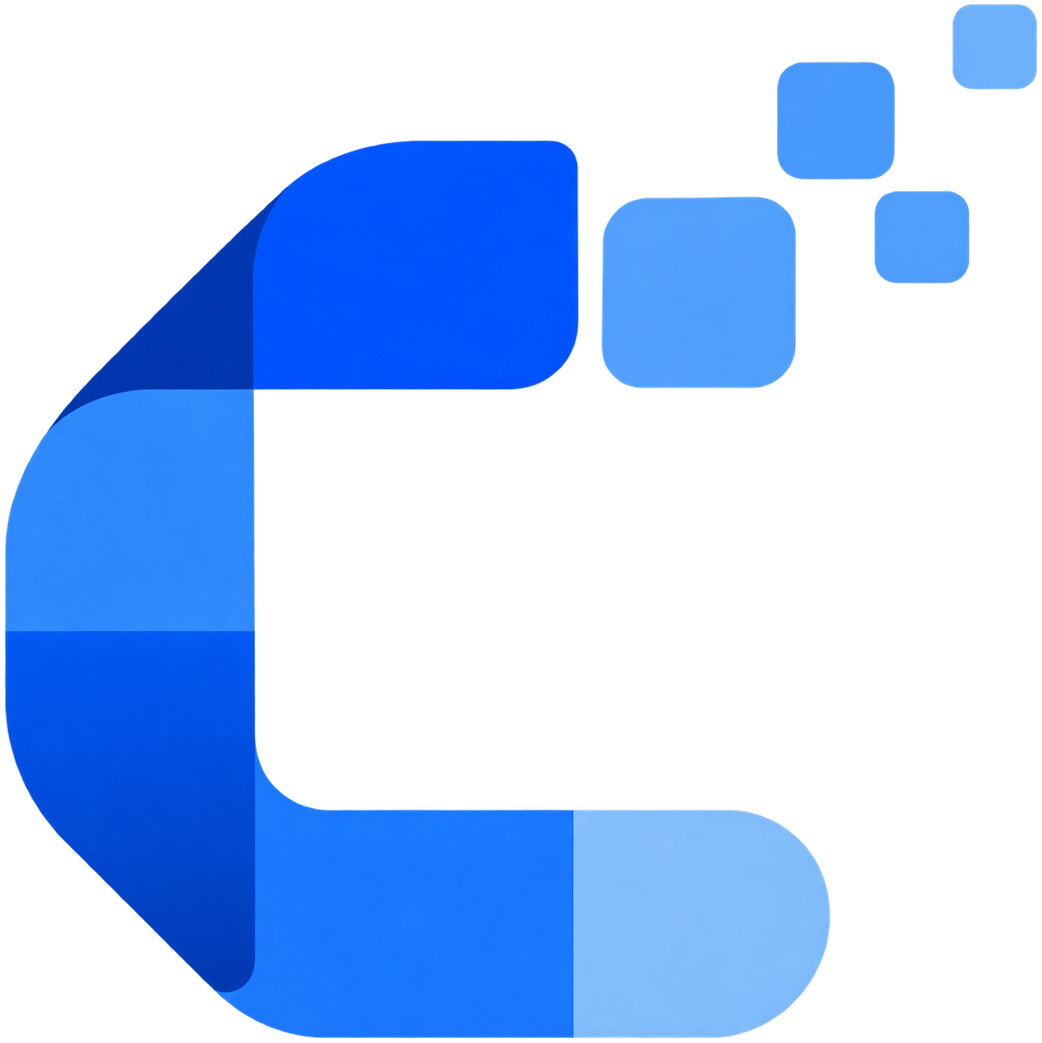}}\hspace{0.3em}Benchmarking Protein Structure Tokenizers\\
       as Generation Targets for Language Models\\[5pt]
       {\fontsize{12}{14}\selectfont Technical Report II: Protein Structure Tokenizers for LLMs}}

\author[1,2]{Yong Liu}
\author[3,4]{Zhanpeng Shi}
\author[2]{Peicong Lin}
\author[1]{Zhenyi Zhong}
\author[5]{Yizhou Dang}
\author[1]{Wendi Wu}
\author[6]{Haoling Li}
\author[2]{Jiayu Gu}
\author[2]{Duo An}
\author[1,2]{Shuangjia Zheng}
\affiliation[1]{Shanghai Jiao Tong University, Shanghai, China}
\affiliation[2]{Lingang Laboratory, Shanghai, China}
\affiliation[3]{Fudan University, Shanghai, China}
\affiliation[4]{Shanghai Innovation Institute, Shanghai, China}
\affiliation[5]{Northeastern University, Shenyang, China}
\affiliation[6]{Nanjing University, Nanjing, China}

\newcommand{\calphaemail}{liuyong9975@sjtu.edu.cn}
\newcommand{\calphahomepage}{https://calpha-ai.com}
\newcommand{\calphagithub}{https://github.com/yliust/Calpha}
\newcommand{\calphahf}{https://huggingface.co/liuyong9975/Calpha}
\newcommand{\linkicon}[1]{\raisebox{-0.14em}{\includegraphics[height=0.95em]{#1}}}
\newcommand{\linkrow}[3]{
  \makebox[1.2em][c]{#1}\enspace\textbf{\sffamily #2} &
  \ifdefempty{#3}{}{\href{#3}{\textcolor{black!65}{\nolinkurl{#3}}}}}
\renewcommand{\checkdatalist}{%
  \begingroup
    \small
    \renewcommand{\UrlFont}{\fontfamily{bytesansmedium}\selectfont}%
    \vspace{1mm}
    \noindent
    \begin{tabularx}{\linewidth}{@{}l@{\hspace{1.8em}}>{\footnotesize\raggedright\arraybackslash}X@{}}
    \makebox[1.2em][c]{\linkicon{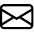}}\enspace\textbf{\sffamily Correspondence} &
      \href{mailto:\calphaemail}{\textcolor{black!65}{\nolinkurl{\calphaemail}}} \\[4pt]
    \linkrow{\linkicon{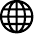}}{Homepage}{\calphahomepage} \\[4pt]
    \linkrow{\linkicon{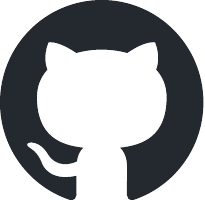}}{GitHub}{\calphagithub} \\[4pt]
    \linkrow{\raisebox{-0.18em}{\includegraphics[height=1.05em]{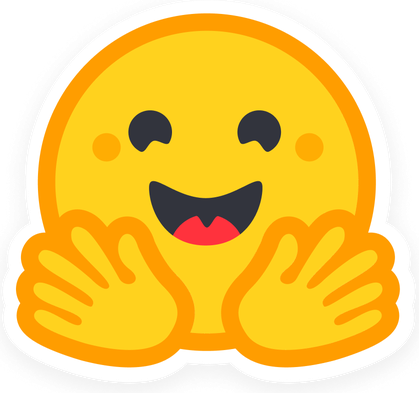}}}{Hugging Face}{\calphahf}
    \end{tabularx}\par
  \endgroup
}

\abstract{%
Discrete structure tokenizers are the standard way to give a language model access to protein
three-dimensional structure, yet they are selected on how accurately they can be \emph{read},
their reconstruction fidelity, while the property that governs generation is how well a model
can \emph{write} them. We measure this property, \emph{writability}, directly: holding the base
model, data, recipe and compute fixed, we swap only the tokenizer and train the same 9B model
to emit structure tokens residue by residue, over eight tokenizers, four compute budgets and
three evaluation sets (96 cells). At 2B tokens final writability runs from TM 0.662 down to
0.378 under identical conditions, with five of the eight arms clearing the TM 0.5 mark at which
a fold is usually called correct and three falling short of it. Neither codebook size nor
reconstruction ceiling explains where an arm lands; the best and worst both use 8192 codewords.
What does explain the spread is local structure preservation of the code space. In DPLM-2 a one-bit codeword error costs
0.019 TM where a random codeword costs 0.347, and the trained model's own errors are
correspondingly local, at 0.51 of the random Hamming expectation; FoldToken4 fails both tests
at the same codebook size and is 0.13 TM less writable. Three controls discipline how such a
benchmark may be read. A same-recipe rerun puts the noise floor at $\sigma=0.0067$ TM, which
collapses an apparent six-way ranking into four tiers. A reference-free geometry panel turns
out to be nearly orthogonal to TM (Pearson $-0.58$), with the worst arm by TM producing the
most physically valid backbones. And two pre-registered interventions are null for a reason the
mechanism measurements reveal: the quantity they target saturates at 1.5B tokens while TM is
still rising. All 131 evaluation cells, including raw generated token streams, are released.
}

\AtBeginDocument{\renewcommand{\bffont}{}\setstretch{1.05}\setlength{\parskip}{4pt plus 1pt minus 1pt}}
\titlespacing*{\section}{0pt}{15pt plus 3pt minus 2pt}{7pt}
\titlespacing*{\subsection}{0pt}{11pt plus 2pt minus 1pt}{5pt}
\titlespacing*{\subsubsection}{0pt}{9pt plus 2pt minus 1pt}{4pt}
\begin{document}

\maketitle
\clearpage

\section{Introduction}

Language models are given access to protein three-dimensional structure by discretising it into
tokens and extending the model's vocabulary with them. A structure tokenizer maps each residue to
one of $K$ codewords, so a backbone becomes a string the model can read alongside text and
sequence. Several such tokenizers are now available off the
shelf~\cite{hayes2025esm3,wang2025dplm2,gao2024foldtoken,yuan2025structtokenbench,gaujac2024language,cao2025protokens},
and a growing number of protein language models are built on
them~\cite{su2024saprot,ma2025prottex,li2024prosst,heinzinger2024prostt5}. This paper studies the
setting in which the model must \emph{emit} that string autoregressively, which turns a structure
tokenizer from a perception module into a generation target. The same shift proved decisive in
other modalities: replacing the tokenizer while holding the language model fixed is what closed
the gap to diffusion in image generation~\cite{yu2024magvit2}, and in audio the codec rather than
the sequence model sets the quality ceiling~\cite{zeghidour2022soundstream,defossez2023encodec}.

The shift changes which property of a tokenizer matters, and that property has gone unmeasured.
Tokenizers are evaluated, and advertised, by reconstruction fidelity: encode a known backbone,
decode it, and report how much structure survives. That number describes the codec. It says how
well structure can be \emph{read} out of the code space, and nothing about whether a model
trained with a next-token objective can \emph{produce} a valid string in it. We call the second
property \emph{writability}. Prior comparisons do not supply it. The systematic benchmark of
structure tokenizers scores reconstruction and frozen-representation
probing~\cite{yuan2025structtokenbench}, both of which are read-side; the closest mechanistic
result, a trade-off between an encoder's locality and its decoder's
reconstruction~\cite{zhang2024aidost}, is measured on the tokenizer alone, with no model asked to
write in the code space. Published systems that do emit structure tokens differ simultaneously in
tokenizer, base model, data and compute, so the tokenizer's own contribution has never been
isolated.

We isolate it. Eight tokenizers are trained into the same 9B base model under one mixture,
recipe, sample order and compute budget, exported at four budgets and scored on three held-out
sets. Three controls then decide what those numbers are allowed to mean. Every score is bracketed
by a floor, uniformly random tokens pushed through that tokenizer's own decoder and scorer, and a
ceiling, ground truth encoded and decoded once; the bracket is necessary because the decoders
differ enough in their structural priors that random tokens alone score anywhere from TM 0.10 to
0.19. The run-to-run noise floor is measured by repeating one arm end to end rather than assumed.
And half the metric panel never looks at the reference, since TM-score~\cite{tmscore},
lDDT~\cite{lddt}, GDT~\cite{zemla2003lga} and RMSD are alike blind to output that is not a
plausible protein, the blindness that motivated physical-validity checks in
docking~\cite{buttenschoen2024posebusters}.

Three findings follow. First, final writability at 2B tokens runs from TM 0.662 (DPLM-2) down to
0.378 (VanillaVQ) under identical conditions, a range that straddles the TM 0.5 mark at which a
fold is usually called correct: five arms clear it and three do not. Neither of the quantities
tokenizers are selected on predicts where an arm lands, since the best and the worst both use
8192 codewords and six of the eight reconstruct above 0.95 TM.
Second, what does predict it is whether nearby codewords denote nearby structures. In DPLM-2,
whose codewords are 13-bit strings, perturbing 30\% of positions by a single bit costs 0.019 TM
while replacing them with random codewords costs 0.347, and the trained model's own errors land
3.29 bits from the correct codeword, about half the random expectation; FoldToken4 has binary
codes and the same codebook size, fails both tests, and is 0.13 TM less writable. A code space is
writable when being slightly wrong is only slightly wrong. Third, the three controls are not
bookkeeping but change the answer: the measured noise floor collapses six apparent ranks into
four tiers, and subtracting the per-tokenizer floor inverts the geometry ranking, because one
decoder returns 81\% valid bond lengths from random input while another returns 3\%.

\section{Related work}
\label{sec:related}

\paragraph{Structure tokenizers are designed and evaluated as codecs.}
Discrete protein structure tokenization has converged on per-residue quantisation of a learned
geometric encoder, differing mainly in the quantiser. ESM3~\cite{hayes2025esm3} and
AminoAseed~\cite{yuan2025structtokenbench} use a learned vector-quantised
codebook~\cite{vandenoord2017vqvae}; DPLM-2~\cite{wang2025dplm2} uses lookup-free binary
quantisation, imported from image tokenization~\cite{yu2024magvit2}; Gaujac et
al.~\cite{gaujac2024language} use finite scalar quantisation~\cite{mentzer2024fsq};
ProSST~\cite{li2024prosst} runs $k$-means over frozen geometric features;
FoldToken4~\cite{gao2024foldtoken} and ProTokens~\cite{cao2025protokens} target compact,
amino-acid-like vocabularies; ProTok~\cite{hardy2026protok} shares one vocabulary across
sequence, structure and function. Foldseek's 3Di alphabet~\cite{vankempen2024foldseek} predates
this line and is designed for search rather than decoding. Across all of them the headline
number is reconstruction, the TM, lDDT or RMSD of a round trip, which measures the codec and not
what a language model can do with it.

\paragraph{The one systematic comparison is read-side, and the closest mechanistic result is
tokenizer-only.}
StructTokenBench~\cite{yuan2025structtokenbench} evaluates structure tokenizers along four
axes: reconstruction, codebook utilisation, and probing of frozen representations on
supervised tasks. It reports that no tokenizer dominates. Every axis asks what can be
\emph{read} out of the code. AIDO.St~\cite{zhang2024aidost} comes closest to our question by
identifying a trade-off between the locality of the encoder and the reconstruction power of
the decoder, but measures it on the tokenizer in isolation; no language model is asked to
write in the code space, and the downstream evidence is representation quality rather than
generation. Our contribution is the write-side measurement these leave open, under a design
that holds the base model, data, recipe and compute fixed.

\paragraph{Systems that emit structure tokens differ in everything at once.}
A number of models extend a language model's vocabulary with structure tokens and generate
them: ESM3~\cite{hayes2025esm3} and DPLM-2~\cite{wang2025dplm2} with iterative masked
decoding in the style of MaskGIT~\cite{chang2022maskgit},
ProtTeX~\cite{ma2025prottex} and SaProt~\cite{su2024saprot} by vocabulary extension of a
protein or general language model, and Gaujac et al.~\cite{gaujac2024language} with a GPT
trained directly on their codes. Each pairs one tokenizer with one base model, mixture and
budget, so the tokenizer's contribution is confounded with every other design choice. This is
precisely the comparison our setup is built to decouple, and it is why we accept the cost of
training eight otherwise-identical arms.

\paragraph{Specialist folding models supply the absolute reference.}
AlphaFold2~\cite{jumper2021af2}, RoseTTAFold~\cite{baek2021rosettafold} and
AlphaFold3~\cite{abramson2024af3} predict structure in continuous space with no discretisation
step, and ESMFold~\cite{lin2023esmfold} does so from a protein language model's features alone.
ESMFold is the right reference for Section~\ref{sec:gap}: it consumes the same ESM2 features
our arms are conditioned on and produces coordinates directly, so the difference isolates the
cost of routing structure through a discrete code and an autoregressive decoder rather than the
cost of a weaker input. CASP~\cite{kryshtafovych2026casp16,yuan2026casp16monomer} and
FoldBench~\cite{xu2025foldbench} supply the held-out targets.

\paragraph{Autoregressive degeneration and reference-free validity are known problems elsewhere.}
Greedy and low-temperature decoding collapse into repetition in natural
language~\cite{holtzman2020degeneration,welleck2020unlikelihood}, and the error accumulation
that teacher-forced training hides from its own metrics has been studied since scheduled
sampling~\cite{bengio2015scheduled,ranzato2016sequence}. The structural-biology analogue of
the second lesson, that a score against a reference cannot see an output that is not a
physically valid molecule, is the argument behind
PoseBusters~\cite{buttenschoen2024posebusters}. Our reference-free panel applies both lessons
to structure tokens: token-level degeneracy statistics on the emitted string, and
geometry checks (bond lengths, clashes, radius of gyration, DSSP
composition~\cite{kabsch1983dssp}) on the decoded backbone.

\section{Measuring writability}
\label{sec:method}

\paragraph{Definition.}
Writability is the structural quality a model attains when it must produce the structure token
string itself: prompted with the amino-acid sequence and a task instruction, it emits one token
per residue by greedy decoding, the string goes through that tokenizer's own decoder, and the
coordinates are scored against the experimental structure with TM-score~\cite{tmscore,zhang2005tmalign}. The
measurement necessarily includes the decoder, which is part of the interface a practitioner
adopts. That inclusion is exactly why the next two controls are mandatory.

\paragraph{Floors and ceilings make scores comparable.}
A raw TM score cannot be compared across tokenizers because each decoder contributes a
different amount of structure on its own. We bracket every model score with two measurements
taken through the identical pipeline on the identical targets: the \emph{floor} $S_4$, from
uniformly random tokens, is what the decoder's prior alone is worth, and the \emph{ceiling}
$S_1$, from encoding and decoding the ground truth once, is the best any model could do in that
code space. We report the \emph{redemption rate}
\begin{equation}
\mathrm{RR} \;=\; \frac{\mathrm{TM}_{\text{model}} - S_4}{S_1 - S_4},
\end{equation}
the fraction of the attainable interval the model actually captured. Figure~\ref{fig:fc}
shows all three quantities together, which is how we report them throughout: a model score
alone is not interpretable.

\paragraph{The redemption rate has a failure mode, and it must be reported with it.}
Normalisation inflates whenever the interval is narrow. Two tokenizers in our set reconstruct
poorly ($S_1 = 0.48$ and $0.42$), and both post redemption rates at the top of the table
(70.9\% and 81.0\%) while being last by raw TM (0.395 and 0.378). A high redemption rate over
a low ceiling means the model has nearly exhausted a code space that cannot express protein
structure in the first place. We therefore never report $\mathrm{RR}$ without $S_1$ beside it,
and we treat these two tokenizers as counterexamples rather than as successes.

\paragraph{A measured noise floor decides how fine a ranking is allowed to be.}
Benchmarks of this kind report strict orderings without knowing the measurement's own variance,
a reporting practice criticised across machine
learning~\cite{reimers2017seeds,henderson2018deeprl,dodge2019showyourwork}.
We estimate it directly: one arm (ESM3) was trained twice end to end under identical recipe,
data and sample order. The nine paired differences have mean $+0.0028$, standard deviation
$0.0067$ and maximum magnitude $0.0123$ TM, four positive and five negative. We adopt
\begin{equation}
|\Delta \mathrm{TM}| < 0.013 \;\Rightarrow\; \text{indistinguishable},
\end{equation}
and apply it to every cross-tokenizer comparison in this paper. This is a lower bound on the
true noise, since it comes from a single repetition of a single arm, so the rule is conservative
in the direction that matters.

\paragraph{Half of the metric panel must not look at the reference.}
TM, lDDT~\cite{lddt}, GDT~\cite{zemla2003lga} and RMSD all compare the output to a target through
the decoder, and all are blind to one class of failure: output that is not a plausible protein at
all---the same gap that physical-validity checks were introduced to close in
docking~\cite{buttenschoen2024posebusters}. We therefore
compute a second panel from the generated structure alone: at the token level, the fraction of
distinct tokens, adjacent-repeat rate, longest run and entropy, which detect the greedy degeneration familiar from
neural text generation~\cite{holtzman2020degeneration,welleck2020unlikelihood};
at the geometry level, the fraction of consecutive CA--CA distances inside
$[3.6, 4.0]$\,\AA{}, the clash rate beyond $|i-j|>2$, and the radius of gyration normalised by
$L^{0.34}$, each between the same floor and ceiling as TM. Because the raw token streams are
archived, this panel was computed retroactively on checkpoints trained weeks earlier, without
rerunning a model.

\section{Experimental setup}
\label{sec:setup}

\paragraph{The controlled variable is the tokenizer and nothing else.}
Every arm starts from the same 9B base model, extends its vocabulary with that tokenizer's $K$
codewords, and trains on the same mixture (structure prediction, inverse folding, infilling,
secondary structure, contacts and general text) built by the same script with the same task
proportions, leakage filtering and sample order. The recipe is identical across arms: one cosine
schedule over 2B tokens, the same learning rates for body, embeddings and adapter, the same
noisy-prefix augmentation, exports at 0.5, 1, 1.5 and 2B tokens. Training runs on
8$\times$H200 at roughly 10 hours per arm. The only deliberate difference is which tokenizer
produced the structure tokens and which decoder scores the output.

\paragraph{Tokenizers.}
We evaluate eight tokenizers spanning codebook sizes from 512 to 8192 and three quantisation
families (Table~\ref{tab:scorecard}): ESM3~\cite{hayes2025esm3,esm3} (learned VQ
codebook~\cite{vandenoord2017vqvae}, $K{=}4096$), DPLM-2~\cite{wang2025dplm2,dplm} (lookup-free binary
quantisation~\cite{yu2024magvit2}, $K{=}8192$), ProFiT~\cite{profit} ($K{=}8192$), FoldToken4~\cite{gao2024foldtoken,foldtoken4} at
levels 10 and 12 ($K{=}1024$ and $4096$, binary codes), AminoAseed~\cite{stb} ($K{=}512$),
GCP-VQVAE~\cite{gcpvq} ($K{=}4096$), and VanillaVQ~\cite{stb} ($K{=}512$). AminoAseed and VanillaVQ share a network and differ only in how the
codebook is trained, which isolates that single factor: VanillaVQ exhibits the codebook
under-utilisation that AminoAseed was designed to
fix~\cite{yuan2025structtokenbench,yu2022vitvqgan}. Repository commits and weight
checksums for all eight are in Appendix~\ref{app:repro}.

\paragraph{Evaluation sets and leakage control.}
We report on three held-out sets: FoldBench334~\cite{xu2025foldbench} (333 chains released after
January 2023), CASP16~\cite{kryshtafovych2026casp16,yuan2026casp16monomer} (155 domains) and
CAMEO2025~\cite{robin2021cameo} (200 single-chain X-ray structures released in 2025).
Leakage control is done when the training set is built, not at evaluation time: a temporal
cut plus sequence-identity filtering~\cite{steinegger2017mmseqs2} against the evaluation targets, with each target's
closest remaining homolog recorded so that results can be split into homologous and
non-homologous buckets. All cells report greedy decoding; sampling and best-of-$n$ are out of
scope for this round and we draw no conclusions about them.

\section{Results}
\label{sec:results}

\begin{figure}[t]
  \centering
  \includegraphics[width=\textwidth]{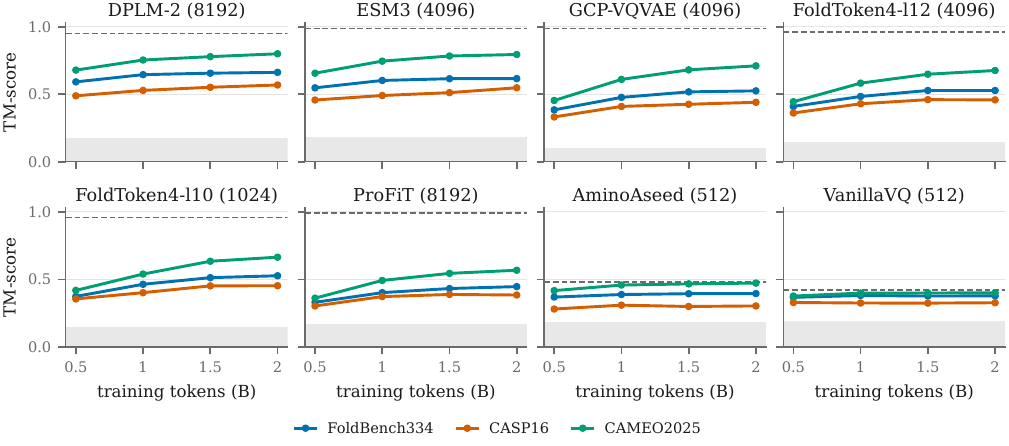}
  \caption{Writability under identical conditions. Each panel is one tokenizer
  (codebook size in parentheses); the three curves are the three evaluation sets.
  The shaded band is that tokenizer's random-token floor $S_4$ and the dashed line is its
  reconstruction ceiling $S_1$, both measured through the same decoder and scorer; panels share
  axes. Identity is carried by panel title rather than colour, since eight categorical hues
  cannot be separated reliably under colour-vision deficiency.}
  \label{fig:curves}
\end{figure}

\subsection{Writability spans TM 0.38 to 0.66, and neither codebook size nor fidelity explains it}

\paragraph{The grid.}
Figure~\ref{fig:curves} shows all 96 cells. At 2B tokens, final TM on FoldBench334 ranges from
0.3781 (VanillaVQ) to 0.6623 (DPLM-2) between arms differing only in tokenizer, and the three
evaluation sets agree on the ordering. We give these as absolute scores and deliberately not as a
ratio between them. TM has no true zero: random tokens already score 0.103 to 0.185 through these
same decoders (Table~\ref{tab:main}), so both ends of the range sit on a pedestal of roughly 0.17,
and a quotient of two raw TM values changes with where the origin is assumed to be. Dividing the
floor-subtracted scores instead would give a different figure again. The quantity with defined
endpoints is the redemption rate of Section~\ref{sec:method}, and that is the normalised number we
report; the ratios that do appear in this paper are between \emph{differences} in TM, which do
have a zero. Except for the two low-ceiling
tokenizers, no arm has saturated; every claim below is a claim at this budget.

\paragraph{Codebook size does not order the arms.}
The best and the worst tokenizer in the table both use 8192 codewords (DPLM-2 at 0.6623,
ProFiT at 0.4467). FoldToken4 at $K{=}1024$ and at $K{=}4096$ land within 0.001 TM of each
other. Size is not the variable.

\paragraph{Reconstruction fidelity does not order them either.}
Six of the eight tokenizers reconstruct above 0.95 TM, and those six span almost the entire
writability range (0.4467 to 0.6623). ProFiT has the highest ceiling in the set ($S_1=0.9902$)
and the lowest redemption rate (33.9\%). A tokenizer can be an excellent codec and a poor
vehicle; reading and writing are different properties.

\subsection{A measured noise floor collapses six ranks into four tiers}

\begin{table}[t]
  \centering
  \caption{Final writability at 2B tokens on FoldBench334, with each tokenizer's own floor and
  ceiling. $\mathrm{RR}$ is the redemption rate. Arms whose TM differs by less than the
  measured noise floor of 0.013 are grouped into one tier and are not ordered within it.
  The two arms marked $\dagger$ have reconstruction ceilings below 0.5: their high
  $\mathrm{RR}$ is an artifact of a narrow interval, not a success.}
  \label{tab:main}
  \begin{tabular}{llrrrrr}
    \toprule
    Tier & Tokenizer & $K$ & $S_4$ (floor) $\downarrow$ & $S_1$ (ceiling) $\uparrow$ & TM $\uparrow$ & RR $\uparrow$ \\
    \midrule
    1 & DPLM-2         & 8192 & 0.177 & 0.952 & \textbf{0.6623} & 62.6\% \\
    2 & ESM3           & 4096 & 0.185 & 0.989 & 0.6162 & 53.7\% \\
    \midrule
    \multirow{3}{*}{3}
      & FoldToken4-l12 & 4096 & 0.144 & 0.962 & 0.5281 & 47.0\% \\
      & FoldToken4-l10 & 1024 & 0.151 & 0.959 & 0.5270 & 46.5\% \\
      & GCP-VQVAE      & 4096 & 0.103 & 0.989 & 0.5254 & 47.7\% \\
    \midrule
    4 & ProFiT         & 8192 & 0.168 & 0.990 & 0.4467 & 33.9\% \\
    \midrule
    \multicolumn{7}{l}{\emph{Degenerate code spaces (reported as counterexamples, not ranked)}}\\
    -- & AminoAseed$^\dagger$ & 512 & 0.185 & 0.482 & 0.3952 & 70.9\% \\
    -- & VanillaVQ$^\dagger$  & 512 & 0.189 & 0.423 & 0.3781 & 81.0\% \\
    \bottomrule
  \end{tabular}
\end{table}

\paragraph{Three of the arms are not distinguishable.}
Table~\ref{tab:main} applies the noise rule from Section~\ref{sec:method}. FoldToken4-l12,
FoldToken4-l10 and GCP-VQVAE are separated by 0.001--0.003 TM, an order of magnitude below the
0.013 noise floor; reporting them in rank order would be reporting noise. The two gaps that do
survive are large: DPLM-2 over ESM3 is 0.046 (3.5$\times$ the noise floor) and ESM3 over
tier 3 is 0.088.

\paragraph{One gap is stable with compute and one is closing.}
With four compute points per arm we can ask whether a gap is a property of the tokenizer or of
the budget. DPLM-2's advantage over ESM3 is flat (0.044, 0.043, 0.041, 0.046); ESM3's over
GCP-VQVAE is not, falling from 0.163 to 0.091. We therefore state the first without
qualification and the second only at $\le$2B tokens. The arms also sit at different points on
their curves (between 1.5B and 2B, ESM3 gains 0.0007 and FoldToken4-l12 $-0.0002$, while ProFiT
and FoldToken4-l10 gain 0.014), so a single-budget snapshot partly reflects convergence rate.

\subsection{Decoder priors differ by an order of magnitude and must be subtracted}

\begin{figure}[t]
  \centering
  \includegraphics[width=\textwidth]{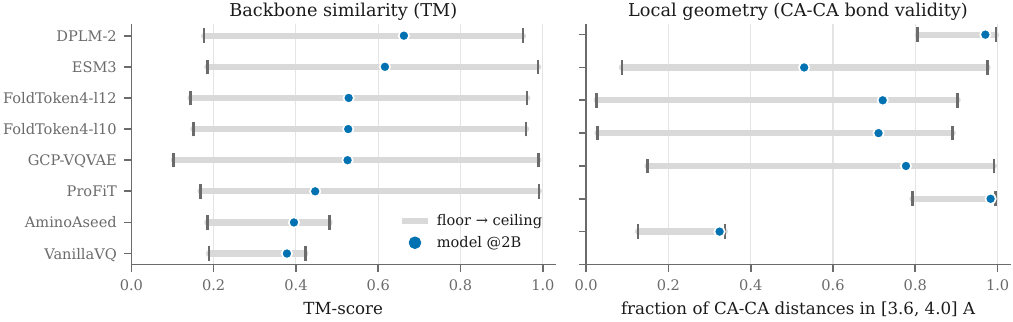}
  \caption{Every score read between its own floor and ceiling, at 2B tokens on FoldBench334.
  Left: backbone similarity. Right: local geometry, the fraction of consecutive CA--CA
  distances inside $[3.6,4.0]$\,\AA{}. Bar ends are the random-token floor and the
  reconstruction ceiling for that tokenizer; the dot is the trained model. The two panels
  order the tokenizers almost inversely.}
  \label{fig:fc}
\end{figure}

\paragraph{Random tokens already produce valid local geometry in some code spaces.}
Passing uniformly random tokens through each decoder yields CA--CA bond-length validity of
0.806 for DPLM-2 and 0.793 for ProFiT, but 0.149 for GCP-VQVAE, 0.087 for ESM3 and 0.026--0.028
for FoldToken4. The decoders differ by an order of magnitude in how much physically plausible
backbone they will emit from no information at all. Any absolute geometry number that is
quoted without this floor attributes the decoder's prior to the model.

\paragraph{After subtracting the floor, the geometry ranking is not the TM ranking.}
DPLM-2's model output is 0.970 valid by raw bond length, near its 0.996 ceiling, but its floor
is 0.806, so it captured 86.3\% of its interval. ProFiT, the weakest arm by TM, reaches 0.983
raw and 94.3\% of its interval, the highest in the set. ESM3, second by TM, is at 0.530 raw and
49.9\% of its interval, with a normalised radius of gyration of 3.33 against 2.79 for
experimental structures. Figure~\ref{fig:fc} shows the two panels side by side.

\paragraph{The two panels are nearly orthogonal.}
Across the seven arms, raw TM and floor-corrected bond validity correlate at Pearson $-0.58$
(Spearman $-0.46$); against the redemption rate, $+0.05$. The reference-free panel is a second
axis, not a redundant check. The extremes make it concrete: ProFiT writes something that looks
like a protein but is not the right protein, while ESM3 writes something whose outline matches
though half its local geometry is invalid and the chain is over-extended.

\paragraph{The degenerate tokenizers fail here too, and in the same way.}
AminoAseed reaches 93.9\% of its bond-validity interval, but its ceiling is 0.337, meaning
ground-truth coordinates passed once through its own codec do not yield valid bond lengths.
This is the normalisation trap of Section~\ref{sec:method} reappearing in a second metric, and
it is the reason we report ceilings in absolute terms everywhere.

\subsection{Corruption tolerance is a property of the representation, not a predictor of writability}

\begin{figure}[t]
  \centering
  \begin{subfigure}[t]{0.32\textwidth}
    \includegraphics[width=\textwidth]{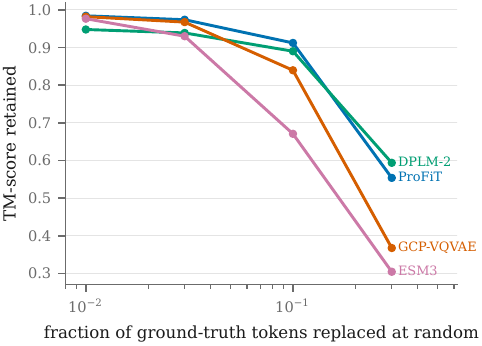}
    \caption{Corruption tolerance.}
    \label{fig:corrupt}
  \end{subfigure}
  \hfill
  \begin{subfigure}[t]{0.32\textwidth}
    \includegraphics[width=\textwidth]{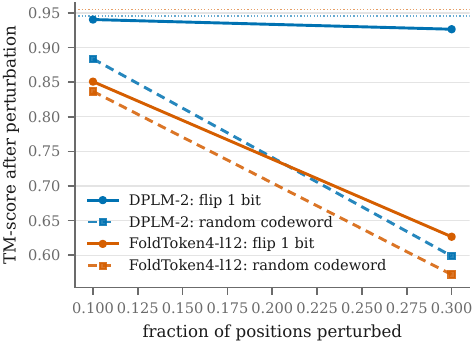}
    \caption{Bit locality.}
    \label{fig:bitloc}
  \end{subfigure}
  \hfill
  \begin{subfigure}[t]{0.32\textwidth}
    \includegraphics[width=\textwidth]{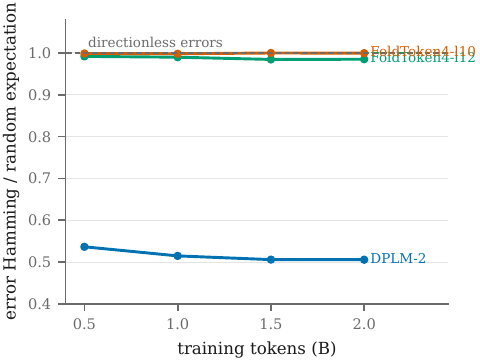}
    \caption{Error geometry over training.}
    \label{fig:hamming}
  \end{subfigure}
  \caption{(a) Replacing a fraction of ground-truth tokens with uniformly random ones;
  this ordering is \emph{not} the writability ordering. (b) Perturbing a fraction of positions
  by one bit (solid) versus by a random codeword (dashed); dotted lines are each tokenizer's
  reconstruction ceiling. (c) Mean Hamming distance of the model's own token errors, divided
  by the expectation for a uniformly random wrong codeword; 1.0 means directionless.}
\end{figure}

\paragraph{Robustness to random token corruption does not predict writability.}
Replacing 10\% of ground-truth tokens at random retains TM 0.912 for ProFiT, 0.890 for DPLM-2,
0.839 for GCP-VQVAE and 0.671 for ESM3 (Figure~\ref{fig:corrupt}): the most corruption-tolerant
tokenizer is the least writable and the least tolerant is the second most writable. Tolerating
\emph{arbitrary} errors is not the same as being easy to write: what matters is whether errors
land on structurally near codewords (Section~\ref{sec:mechanism}) and whether long-range
contacts survive (Section~\ref{sec:gap}). We report the disagreement rather than select the
metric that agrees.

\section{What makes a code space writable}
\label{sec:mechanism}

\paragraph{The hypothesis.}
The previous section rules out codebook size, reconstruction fidelity and corruption tolerance.
The candidate left standing is a property of the code space's local structure: a tokenizer is
writable if
codewords that are close in the model's output space denote structures that are close in
Cartesian space, so that a near-miss is a near-hit. For tokenizers whose codewords are binary
codes this is directly testable, because "close in output space" has a concrete meaning: a
small Hamming distance~\cite{hamming1950}. This is the property an error-correcting output code
buys deliberately~\cite{dietterich1995ecoc}; here it is a side effect of the quantiser, which is
why it varies so much across tokenizers. The encoder-side locality that AIDO.St trades against
reconstruction~\cite{zhang2024aidost} is the same quantity seen from the read side. Two of our tokenizers qualify: DPLM-2, whose lookup-free quantiser~\cite{yu2024magvit2}
makes each codeword a 13-bit string, and FoldToken4, whose level-$\ell$ codes are
$\ell$-bit strings.

\paragraph{Test 1: do neighbouring codewords denote neighbouring structures?}
We encode ground-truth backbones, perturb a fraction of positions, and decode
(Figure~\ref{fig:bitloc}). Perturbing by \emph{one bit} should be cheap if the code is locally
structure-preserving; perturbing by a \emph{random codeword} is the control. For DPLM-2 at
30\% of positions, the one-bit perturbation costs 0.019 TM (0.946 to 0.927) while the random
codeword costs 0.347 (to 0.599), a factor of 18. For FoldToken4-l12 the two perturbations
cost almost the same (0.851 versus 0.837 at 10\%, 0.627 versus 0.572 at 30\%): its bits do not
encode structural similarity, despite being bits.

\paragraph{Test 2: are the model's own errors local?}
Test 1 is a property of the tokenizer. The question that matters is whether the trained model
exploits it. We take the archived greedy generations and measure, for every token the model got
wrong, the Hamming distance between what it wrote and what was correct, against the expectation
for a uniformly random wrong codeword (Figure~\ref{fig:hamming}). At 2B tokens DPLM-2's model
errs by 3.29 of 13 bits, a ratio of 0.506; FoldToken4-l12 is at 0.985 and FoldToken4-l10 at
0.999, indistinguishable from writing a random wrong codeword. The tokenizer whose bits carry
structure is the tokenizer whose model makes structurally cheap mistakes, and it is the
tokenizer that is 0.13 TM more writable.

\paragraph{The quantity saturates well before training ends.}
Because the generations are archived at all four budgets, this is a trajectory rather than a
point. DPLM-2's ratio falls 0.537, 0.515, 0.506, 0.506 over 0.5--2B tokens, stopping at 1.5B,
while TM is still rising (0.5919, 0.6455, 0.6564, 0.6623): whatever drives the last quarter of
the TM gain, it is not further locality of the model's errors. FoldToken4's two levels sit at
0.98--1.00 throughout, unmoved by 2B tokens of training.

\paragraph{What this does and does not establish.}
The locality account is correlational across eight tokenizers plus two direct measurements of
the mechanism itself, and is the only candidate we tested that survives. It is not established
causally: the intervention designed to test it (Section~\ref{sec:negative}) is null, with the
saturation result above as the explanation. We state it as the best-supported explanation of
the ranking, not as a demonstrated cause.

\section{Where the gap to a specialist folding model goes}
\label{sec:gap}

\paragraph{The absolute number, not the relative one.}
The previous two sections compare tokenizers to each other and to their own ceilings. Neither
asks why the best arm reaches TM 0.662 when ESMFold~\cite{lin2023esmfold}, on the \emph{same} 333
targets from the same ESM2-3B features, reaches 0.870. ESMFold is the appropriate reference
because it consumes those same features and emits coordinates directly, with no discrete code in
between, so the difference isolates the cost of the token interface rather than of a weaker
input. The paired difference is $+0.209$ (bootstrap 95\% CI~\cite{efron1979bootstrap}
$[+0.194, +0.227]$) and ESMFold wins on 96.3\% of targets, so the shortfall is systematic. This
section decomposes it. The representation is not the constraint: DPLM-2 reconstructs at TM 0.952
and recovers 0.983 of long-range contacts from ground-truth tokens, both above 0.870, so what is
lost is lost between the input and the emitted string.

\paragraph{Long-range contacts are where tokenizers differ, and where this one falls short.}
We split contacts by sequence separation and score each band through the same pipeline, between a
floor (nearest-in-sequence prediction) and a ceiling (ground-truth tokens decoded). DPLM-2
captures 72.3\% of attainable short-range contacts ($6 \le |i-j| < 12$), 71.3\% of medium and
63.2\% of long-range ($|i-j| \ge 24$). A 9-point gradient within one arm looks minor; across arms
it is the dominant axis (Table~\ref{tab:bands}). Long-range capture spans 63.2\% to 20.3\%, a
42.8-point range against 15.4 for short range, and correlates with final TM at $+0.976$ versus
$+0.786$. \textbf{What separates a writable tokenizer from an unwritable one is whether
long-range contacts survive into the generated string.} This also explains the inversion of
Section~\ref{sec:results}: ProFiT produces the most valid backbones and the worst TM because it
writes locally plausible fragments with the wrong global topology.

\begin{table}[t]
  \centering
  \caption{Contact capture by sequence separation at 2B tokens on FoldBench334, as the fraction
  of the floor-to-ceiling interval the model realises. Floors (nearest-in-sequence) are 0.071 /
  0.047 / 0.033 and ceilings (ground-truth tokens decoded) are 0.91--0.94 / 0.90--0.93 /
  0.98--0.99 across arms. The long-range band spreads the arms nearly three times as far as the
  short-range band.}
  \label{tab:bands}
  \begin{tabular}{lrrrr}
    \toprule
    Tokenizer & TM @2B $\uparrow$ & short $\uparrow$ & medium $\uparrow$ & long $\uparrow$ \\
    \midrule
    DPLM-2         & 0.6623 & 72.3\% & 71.3\% & \textbf{63.2\%} \\
    ESM3           & 0.6162 & 67.6\% & 67.0\% & 58.7\% \\
    FoldToken4-l12 & 0.5281 & 67.9\% & 63.8\% & 44.6\% \\
    GCP-VQVAE      & 0.5254 & 56.9\% & 52.1\% & 35.4\% \\
    ProFiT         & 0.4467 & 59.5\% & 49.2\% & \textbf{20.3\%} \\
    \midrule
    \multicolumn{2}{l}{spread across arms} & 15.4\,pt & 22.1\,pt & \textbf{42.8\,pt} \\
    \multicolumn{2}{l}{correlation with TM} & $+0.786$ & $+0.906$ & $\mathbf{+0.976}$ \\
    \bottomrule
  \end{tabular}
\end{table}

\paragraph{Half the gap is long-range contacts the model fails to transfer.}
The input already contains more long-range information than the model emits: ESM2-3B's own
unsupervised contact head~\cite{rao2021contacts} reaches P@L/5 $=0.813$ on these targets, against the model's 0.633. Regressing
per-target TM on per-target long-range precision gives a slope of $0.505$ for DPLM-2, and
essentially the same slope for every other arm (0.526, 0.530, 0.529; $r = 0.83$--$0.87$), so the
exchange rate between contact precision and backbone similarity is a property of the pipeline
rather than of one tokenizer. Moving the model from its 0.633 to the input's 0.813 therefore
corresponds to $+0.091$ TM, which is \textbf{44\% of the 0.209 gap}. The per-target split agrees:
on the third of targets where the model's long-range precision is highest it trails ESMFold by
0.119, and on the worst third by 0.325.

\paragraph{The other half is the cost of discretisation itself.}
Extrapolating the same regression to the representation's own ceiling (long-range 0.983) predicts
TM 0.840, still below ESMFold's 0.870. Even a model that transferred every long-range contact
its code space can express would not close the gap. Consistently, on the third of targets where
the model's long-range contacts are best, it still trails by 0.119. We therefore report the
decomposition as roughly half long-range transfer and half an irreducible discretisation cost,
with the caveat that the second term is obtained by extrapolation beyond the observed range.

\paragraph{There is no sparse set of critical errors.}
If a few catastrophic tokens dominated, confidence-based repair or local resampling would recover
most of the loss. Starting from the model's own greedy tokens we replaced a random fraction $x$
with ground truth and re-decoded; recovery is \emph{sub-linear}. For DPLM-2, $x=10\%$ recovers
5\% of the distance to the ceiling, $x=20\%$ recovers 12\% and $x=40\%$ recovers 28\%; ESM3 is
more extreme at 3\%, 7\% and 18\%. Both curves are convex: the marginal value of fixing a token
\emph{increases} with how many are already fixed, so errors must be corrected in bulk, as
expected if topology is determined cooperatively. Our pre-registered threshold would have called
the errors sparse had $x=10\%$ recovered half the gap; it recovers a twentieth.

\paragraph{Better decoding is a small lever, and we measured how small.}
If the distribution held much better structures than greedy selects, better decoding would be the
cheapest remedy. We sampled 32 chains per target at $T{=}1.0$ and scored all of them, reporting
the oracle maximum (best by TM, an upper bound that uses the answer), the medoid (the sample most
similar to the others, which uses no reference and is therefore available and is the structural
analogue of self-consistency selection~\cite{wang2023selfconsistency}), and the maximum over 32
random token strings, which extreme-value statistics require as a matched draw.
The oracle maximum is $+0.0515$ TM over greedy (paired 95\% CI $[+0.0448, +0.0589]$, better on
84.4\% of targets), confirming that better structures exist in the distribution. The medoid
recovers only $+0.0095$ ($[+0.0036, +0.0157]$, 53.2\% of targets): \textbf{18\% of the oracle
gain is reachable without the reference}. The random-token control sits at 0.2208, far below
either, so the effect is not an artifact of sampling many times. Decoding and selection are
therefore worth about $0.01$ TM in practice, roughly 5\% of the gap, at 32 times the inference
cost, against an upper bound of 25\% that is not attainable.

\paragraph{Summary of the decomposition.}
Table~\ref{tab:gap} collects the terms. Long-range transfer dominates, an irreducible
discretisation cost is second, and decoding is a minor term. The three are not strictly
additive, since improving long-range transfer would change what is left for the other two, so the
percentages are magnitudes rather than an exact partition. The practical reading is that neither
a better decoder nor a better training objective alone closes this gap: the first is worth a
hundredth of a TM point, and the second runs into a discretisation floor around TM 0.84.

\begin{table}[t]
  \centering
  \caption{Decomposition of the 0.209 TM gap between the most writable arm (DPLM-2 at 2B tokens)
  and ESMFold on the same 333 targets from the same ESM2-3B features. Terms overlap and are not
  strictly additive.}
  \label{tab:gap}
  \begin{tabular}{llrl}
    \toprule
    Term & Magnitude & Share & Evidence \\
    \midrule
    Long-range contacts not transferred & $\approx 0.091$ & 44\% & input 0.813 vs model 0.633; slope 0.505--0.530 \\
    Discretisation cost                 & $\approx 0.07$  & 33\% & ceiling 0.983 predicts TM 0.840 $<$ 0.870 \\
    Decoding / selection (attainable)   & $0.0095$        & 5\%  & medoid vs greedy, paired \\
    \quad (oracle upper bound)          & $(0.0515)$      & (25\%) & best-of-32 with the reference \\
    \bottomrule
  \end{tabular}
\end{table}

\subsection{Where the pipeline fails outright: chains beyond 500 residues}
\label{sec:length}

\paragraph{The length effect is not what it is usually assumed to be.}
Long chains are commonly expected to degrade monotonically, and for the most writable arm they do
not. DPLM-2's per-target TM correlates with chain length at only $-0.054$, and its best bucket is
300--500 residues (TM 0.726), above both 100--200 (0.638) and 200--300 (0.674).
Below 500 residues there is no length penalty worth reporting.

\begin{figure}[t]
  \centering
  \includegraphics[width=\textwidth]{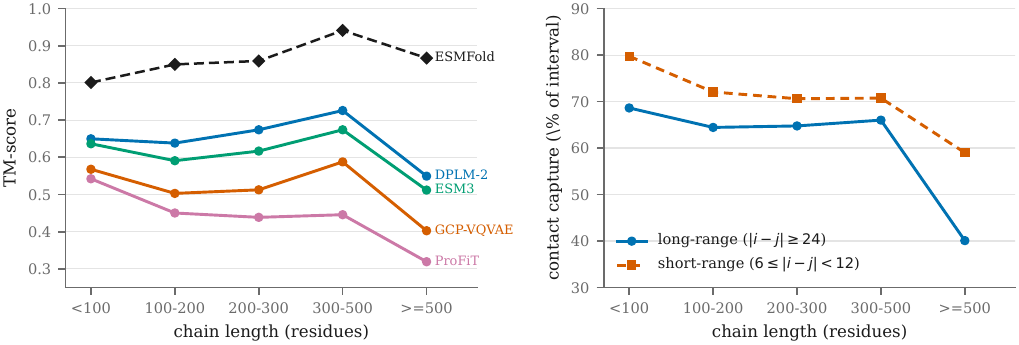}
  \caption{Chain length. Left: TM by length bucket for four arms and ESMFold on the same targets.
  Model arms peak at 300--500 residues and fall beyond 500; ESMFold rises with length, so the gap
  widens monotonically. Right: contact capture by band for DPLM-2, as the fraction of the
  floor-to-ceiling interval realised. The long-range band collapses beyond 500 residues while the
  short-range band degrades gently.}
  \label{fig:length}
\end{figure}

\paragraph{There is a cliff at 500 residues, and it is a long-range failure.}
Beyond 500 residues TM falls to 0.549, and the reference-free decomposition says why: long-range
contact capture collapses from 66.0\% in the 300--500 bucket to 40.1\%, while short-range capture
falls much less, from 70.7\% to 59.0\% (Figure~\ref{fig:length}, right). The bucket holds 26
targets (25 of which also have a contact panel, from which the capture figures are computed), so
the point estimate is loose; but the gap to ESMFold in that bucket ($0.327$, paired 95\%
interval $[+0.254, +0.390]$) does not overlap any other bucket's interval, the widest of which
reaches $+0.248$.
This is the same mechanism as Section~\ref{sec:gap} appearing as an operating regime: long chains
contain proportionally more long-range contacts, and long-range contacts are what this pipeline
fails to transfer.

\paragraph{The gap to a specialist model widens with length because the specialist improves.}
The model's own degradation with length is mild, yet the gap to ESMFold grows monotonically from
$0.151$ (under 100 residues) to $0.327$ (500 and above), because ESMFold \emph{gains} with length
(TM correlates $+0.172$ with chain length, against $-0.054$ for DPLM-2). Longer chains tend to
carry more evolutionary signal and more regular domain structure, and a specialist folding model
converts that into accuracy while the token pipeline does not. Reporting only the mean over a
length-mixed benchmark hides this: the two systems are not falling behind at the same rate in the
same places.

\paragraph{Length has an effect of its own, beyond long-range contacts.}
Chain length and long-range capture are negatively correlated ($-0.183$), so part of the length
effect is the contact effect. It is not all of it: after regressing the per-target gap on
long-range capture, the residual still correlates with length at $+0.195$. Something that scales
with length, whether the number of autoregressive steps, accumulated commitment or attention
span, costs accuracy independently of which contacts are captured. Error accumulation over a longer
generation is the behaviour exposure-bias analyses would
predict~\cite{bengio2015scheduled,ranzato2016sequence}. We have not isolated that term, and we do
not claim a decomposition of it.

\paragraph{Length sensitivity is itself a writability signal.}
The correlation between TM and chain length is $-0.054$ for DPLM-2, $-0.072$ for ESM3,
$-0.102$ for GCP-VQVAE, $-0.167$ and $-0.187$ for FoldToken4-l10 and l12, and $-0.264$ for
ProFiT, the same ordering as writability itself (Table~\ref{tab:main}). A tokenizer whose code
space transfers long-range structure poorly is also a tokenizer whose quality decays with length,
which is the behaviour one would predict from Section~\ref{sec:mechanism}.

\paragraph{Scope.}
Generation is capped at 1024 structure tokens. Exactly one of the 334 evaluation chains is
longer than that and is excluded, so the cliff at 500 residues is well inside the generated range
and is not an artifact of the cap. All numbers in this subsection are for
greedy decoding at the 2B-token budget on FoldBench334; we did not test whether the cliff moves
with compute, which would need the length buckets re-evaluated at every checkpoint.

\section{Two pre-registered negative results}
\label{sec:negative}

\subsection{Supervising the bits directly changes nothing}

\paragraph{The intervention.}
If writability comes from the model landing on bit-adjacent codewords, then training the model
to predict the \emph{bits} rather than the codeword index should help. We pre-registered two
implementations and their decision thresholds before running them. The first (D2H) keeps the
index cross-entropy but smooths the target over Hamming-near codewords with
$\epsilon = 0.2$. The second (D2B) replaces the index objective with a per-bit binary
cross-entropy on the bit marginals, which is the objective an error-correcting output code
prescribes~\cite{dietterich1995ecoc}. Both use DPLM-2 tokens and are otherwise identical to the
DPLM-2 arm.

\paragraph{Both are null.}
D2H was trained to 2B tokens and evaluated on all twelve cells: every difference from the
baseline lies within $\pm 0.016$ TM, and the pre-registered primary comparison
(FoldBench334 at 2B) is $-0.0001$ with a paired bootstrap interval~\cite{efron1979bootstrap} of
$[-0.0078, +0.0075]$ and a 50.8\% win rate. D2B reached 1.39B tokens before being stopped;
at its 0.5B checkpoint the three evaluation sets give $+0.002$, $+0.014$ and $-0.002$.
No cell moves outside the noise.

\paragraph{The pre-registered mechanism check is what makes the null interpretable.}
The same plan required that a null result be discarded unless the intervention demonstrably
changed the error geometry. For D2H it did not: the error Hamming ratio is 0.507 against the
baseline's 0.506, and the error rate is unchanged at 92.3\% versus 92.4\%. That run therefore
tested nothing and we report it as void rather than as evidence against bit supervision;
$\epsilon = 0.2$ was simply too gentle a label smoothing for the gradient to reach the bits.
D2B did change the mechanism, pushing the ratio from 0.537 to 0.526 at 0.5B, so its null is
informative. The explanation is the saturation result of Section~\ref{sec:mechanism}: the
quantity these interventions target stops improving at 1.5B tokens under ordinary index
supervision, so there was little left to teach.

\paragraph{A by-product worth more than the intervention.}
D2B at 0.5B tokens has a \emph{higher} token error rate than the baseline (95.1\% versus
93.9\%) and the same TM. Token accuracy and structural quality are decoupled: which codeword
the model writes matters much less than how far it lands from the right one. Any evaluation of
structure tokenizers that uses token accuracy as a proxy for structural quality is measuring
the wrong thing.

\subsection{Forcing a correct prefix does not rescue the worst tokenizer}

\paragraph{The hypothesis being tested.}
ProFiT has near-zero prefix consistency ($S_5 = 0.0011$ against ESM3's 0.5927, measured at a
half-length prefix on principal-axis-canonicalised inputs to match the training condition):
re-encoding a truncated chain changes almost every token. A natural explanation of its poor writability is
that autoregressive generation must commit to early tokens that cannot be determined from a
prefix. If that were the mechanism, replacing the prefix with ground truth should recover
much of the loss.

\paragraph{The measurement must exclude the gift.}
We force the first $q \in \{0.25, 0.5, 0.75\}$ of the tokens to ground truth and score
\emph{only the residues the model wrote itself}, superposing on that segment alone and
comparing against the same segment under ordinary greedy decoding, with a paired bootstrap.
Scoring the whole chain would count the copied prefix as model performance: at $q=0.25$ that
inflates the ESM3 effect from $+0.029$ to $+0.078$, and the inflation grows with $q$.

\begin{figure}[t]
  \centering
  \includegraphics[width=\textwidth]{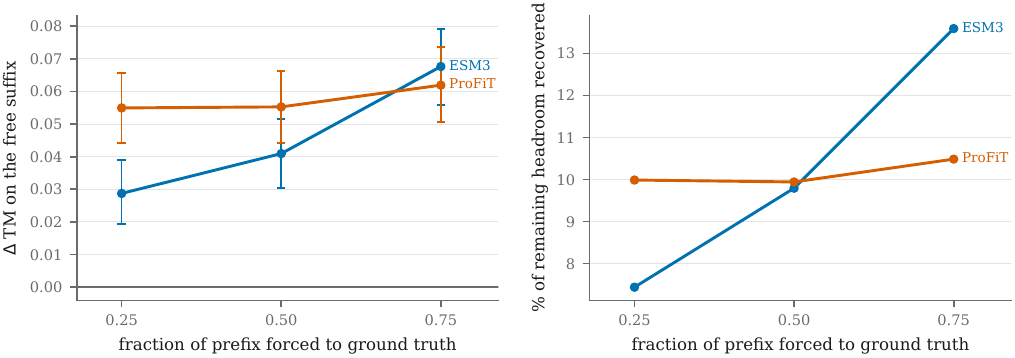}
  \caption{Ground-truth prefix rescue, scored only on the suffix the model wrote itself.
  Left: absolute $\Delta$ TM with paired bootstrap 95\% intervals. Right: the same effect as a
  fraction of the headroom remaining to each tokenizer's reconstruction ceiling, which removes
  the advantage ProFiT gets from starting lower.}
  \label{fig:rescue}
\end{figure}

\paragraph{ProFiT does not respond; ESM3 does.}
Figure~\ref{fig:rescue} gives the result. ProFiT's gain is flat across the three settings
($+0.055$, $+0.055$, $+0.062$ TM; 10.0\%, 9.9\%, 10.5\% of the remaining headroom to its
ceiling), while ESM3 rises monotonically ($+0.029$, $+0.041$, $+0.068$; 7.4\%, 9.8\%, 13.6\%)
and overtakes ProFiT at $q=0.75$. Handing ProFiT three quarters of the answer helps no more
than handing it one quarter. All six intervals exclude zero, so the effect is real and small;
what fails is the hypothesis, not the measurement.

\paragraph{We withdraw the causal reading of prefix consistency.}
Low $S_5$ is a symptom that co-occurs with low writability, not its cause; the bottleneck for
ProFiT is the representation itself (Section~\ref{sec:mechanism}), not the commitment that
autoregressive decoding forces. An independent measurement of where along the chain errors
occur had already failed to support the commitment hypothesis, and this intervention agrees.
In particular, these data do not support recommending bidirectional or diffusion
decoding~\cite{chang2022maskgit,wang2024dplm} as a remedy, a recommendation that would have
followed from the causal reading.

\section{General capability is retained, and training loss cannot tell you so}
\label{sec:guardrail}

\paragraph{The guardrail.}
Extending the vocabulary and continuing pretraining could simply damage the base model through
catastrophic forgetting~\cite{kirkpatrick2017forgetting,luo2023forgetting}, a known hazard of
domain-adaptive pretraining~\cite{gururangan2020dapt}, which would be an alternative explanation
for any writability difference. We evaluate all seven
2B-token checkpoints on a pre-registered 350-item stratified subset of MMLU-Pro~\cite{wang2024mmlupro,mmlupro}, using the
upstream project's item list and scorer unchanged so that the numbers are comparable with
earlier rounds. The base model scores 0.7743; extending the vocabulary without training leaves
it at 0.7743, item for item.

\paragraph{No arm collapses.}
Accuracies range from 0.7629 (ESM3, $-1.14$pp) to 0.7257 (ProFiT, $-4.86$pp), with parse
failure rates between 0.00\% and 0.86\% against the base model's 0.29\%
(Table~\ref{tab:mmlu}); nothing resembles format collapse, so the alternative explanation is
ruled out.

\begin{figure}[t]
  \centering
  \includegraphics[width=\textwidth]{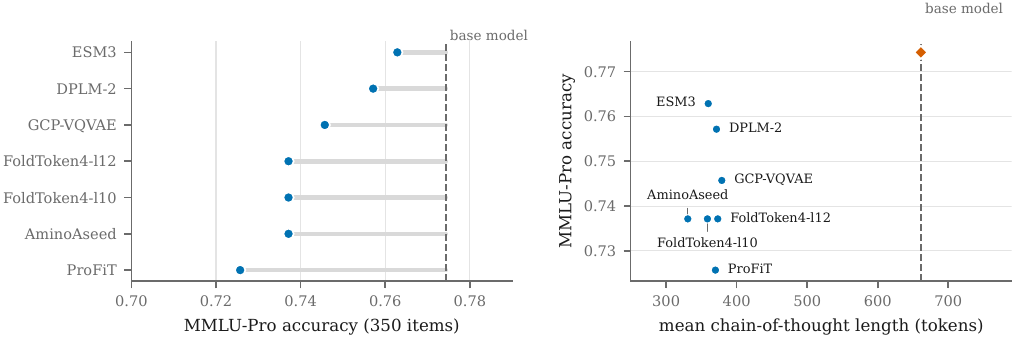}
  \caption{General-capability guardrail at 2B tokens. Left: MMLU-Pro accuracy against the base
  model (dashed). Right: accuracy versus mean chain-of-thought length; the diamond is the base
  model. Every arm loses roughly half its reasoning length regardless of tokenizer.}
  \label{fig:mmlu}
\end{figure}

\paragraph{The model does not get less capable; it gets terser.}
Mean chain-of-thought~\cite{wei2022cot} length falls from 661 tokens for the base model to 331--379 for every
arm, a 43--50\% reduction with no systematic relation to the tokenizer
(Figure~\ref{fig:mmlu}). Because all seven arms behave the same way, this is a property of the
training recipe, not of the structure tokens. Reporting accuracy alone would have hidden it
entirely, for the same reason that reporting TM alone hides the failures of
Section~\ref{sec:results}.

\paragraph{The in-mixture text loss is blind to this.}
The final general-text training loss across the seven arms is 2.1663, 2.1673, 2.1682, 2.1691,
2.1692, 2.1694 and 2.1695, a spread of 0.003, while their MMLU-Pro accuracies differ by 3.7pp.
Training loss on in-mixture text measures fit, not retention; used as a guardrail it would
have reported that all seven arms were equally safe.

\paragraph{We do not claim that more writable arms retain more.}
Raw TM and accuracy drop correlate at Pearson $+0.825$ (permutation $p=0.026$), but no pairwise
comparison is significant: the most extreme pair, ESM3 versus ProFiT, differing by 3.7pp, gives
an exact McNemar test~\cite{mcnemar1947} $p = 0.079$, and three of seven arms score identically (258/350). A 350-item
set has a binomial standard error of about 2.3pp. We report the association as an observation
and not as a finding; settling it needs the full 12{,}032-item benchmark.

\section{Limitations and what we do not claim}
\label{sec:limits}

We state the boundary of every claim explicitly, because the controls in
Section~\ref{sec:method} exist precisely to stop this benchmark from being over-read.

\textbf{We do not claim a best tokenizer.} We claim that DPLM-2 is the most writable of the
eight under this base model, this recipe and a 2B-token budget. No arm except the two
degenerate ones has saturated.

\textbf{We do not claim that ESM3 beats GCP-VQVAE.} That gap closes from 0.163 to 0.091 as
compute grows and would plausibly vanish beyond our budget; compute-matched comparison is the
only kind that is meaningful here~\cite{kaplan2020scaling,hoffmann2022chinchilla}. It is reported as a statement
about $\le$2B tokens only.

\textbf{We do not order the third tier}, whose three members differ by less than the noise floor,
and \textbf{we do not treat a high redemption rate as a success on its own}: for AminoAseed and
VanillaVQ it reflects a narrow floor-to-ceiling interval.

\textbf{We do not claim bit locality is causal.} It is the best-supported correlate, with two
independent mechanism measurements; the intervention that would establish causality is null
(Section~\ref{sec:negative}), and we report that null and its explanation rather than omit it.

\textbf{We do not claim writable arms retain more general capability}, since the association is
not significant pairwise, and \textbf{we do not generalise beyond one configuration}: one base
model, one scale, one mixture, one recipe, greedy decoding only, with no result about sampling
or best-of-$n$.

\textbf{We do not claim a general length penalty.} Below 500 residues the most writable arm shows
no length effect worth reporting ($r=-0.054$); the failure is specific to chains of 500 residues
and above, where only 26 targets are available (Section~\ref{sec:length}). We also did not
isolate the length term that survives conditioning on long-range capture, and we did not test
whether the 500-residue cliff moves with compute.

\textbf{The noise floor is a lower bound}, from one repetition of one arm ($n=9$ paired cells);
three to five repetitions per arm, at 30--50 GPU-hours each, would be the proper estimate.
\textbf{The reference-free panel is partial}: secondary-structure composition needs a full
backbone, so for CA-only decoders we report bond validity, clash rate and radius of gyration
but not DSSP composition. \textbf{The gap decomposition's second term is an extrapolation}
beyond the observed range of long-range precision, and its three terms overlap.

\section{Conclusion}

Structure tokenizers are chosen by how well they can be read, and used as something that must be
written. Measuring the second property directly, with everything else fixed, we find it varies by
from TM 0.662 to 0.378 across eight tokenizers, the distance between a backbone a reader would
call the right fold and one they would not, and is predicted by neither codebook size nor
reconstruction fidelity, but by whether the code space is locally structure-preserving, so that
the model's inevitable mistakes are structurally cheap.

For practice, this implies three changes. When selecting a tokenizer for a generative interface, measure
what a one-codeword error costs, not only what a perfect round trip preserves. When reporting
through a generative decoder, bracket every number with a random-input floor and a ground-truth
ceiling on the same targets, and give the ceiling in absolute terms so that normalised scores
over narrow intervals are not mistaken for success. And never use token accuracy as a proxy for
structural quality: we exhibit a variant that raises the token error rate while leaving
structural quality unchanged.

Our decomposition of the gap to a specialist folding model also bounds what the remaining
levers are worth: better decoding buys about 0.01 TM in practice, transferring the long-range
contacts already present in the input would buy 0.09, and a discretisation floor near TM 0.84
sits behind both. Closing the remaining distance therefore requires either a better-quantised code
space~\cite{mentzer2024fsq,yu2024magvit2,zhang2024aidost} or a continuous refinement
stage~\cite{abramson2024af3}, not another decoding heuristic.

\bibliographystyle{oridynamic_odesign}
{\setlength{\bibsep}{3pt plus 0.3pt}\bibliography{references}}

@article{jumper2021af2,
  title   = {Highly accurate protein structure prediction with {AlphaFold}},
  author  = {Jumper, John and Evans, Richard and Pritzel, Alexander and others},
  journal = {Nature}, volume = {596}, pages = {583--589}, year = {2021}
}

@article{abramson2024af3,
  title   = {Accurate structure prediction of biomolecular interactions with {AlphaFold} 3},
  author  = {Abramson, Josh and Adler, Jonas and Dunger, Jack and others},
  journal = {Nature}, volume = {630}, pages = {493--500}, year = {2024}
}

@article{lin2023esmfold,
  title   = {Evolutionary-scale prediction of atomic-level protein structure with a language model},
  author  = {Lin, Zeming and Akin, Halil and Rao, Roshan and others},
  journal = {Science}, volume = {379}, number = {6637}, pages = {1123--1130}, year = {2023}
}

@article{hayes2025esm3,
  title   = {Simulating 500 million years of evolution with a language model},
  author  = {Hayes, Thomas and Rao, Roshan and Akin, Halil and others},
  journal = {Science}, volume = {387}, number = {6736}, pages = {850--858}, year = {2025}
}

@inproceedings{wang2025dplm2,
  title     = {{DPLM-2}: A Multimodal Diffusion Protein Language Model},
  author    = {Wang, Xinyou and Zheng, Zaixiang and Ye, Fei and Xue, Dongyu and Huang, Shujian and Gu, Quanquan},
  booktitle = {International Conference on Learning Representations},
  year      = {2025}
}

@article{tmscore,
  title   = {Scoring function for automated assessment of protein structure template quality},
  author  = {Zhang, Yang and Skolnick, Jeffrey},
  journal = {Proteins: Structure, Function, and Bioinformatics},
  volume  = {57}, number = {4}, pages = {702--710}, year = {2004}
}

@article{lddt,
  title   = {lDDT: a local superposition-free score for comparing protein structures and
             models using distance difference tests},
  author  = {Mariani, Valerio and Biasini, Marco and Barbato, Alessandro and Schwede, Torsten},
  journal = {Bioinformatics}, volume = {29}, number = {21}, pages = {2722--2728}, year = {2013}
}

@misc{mmlupro,
  key          = {MMLU-Pro},
  title        = {{MMLU-Pro}: A More Robust and Challenging Multi-Task Language Understanding
                  Benchmark},
  howpublished = {\url{https://huggingface.co/datasets/TIGER-Lab/MMLU-Pro}},
  note         = {Evaluated on a pre-registered 350-item stratified subset with the upstream
                  item list and scorer}
}

@article{baek2021rosettafold,
  author  = {Baek, Minkyung and DiMaio, Frank and Anishchenko, Ivan and others},
  title   = {Accurate prediction of protein structures and interactions using a three-track neural network},
  journal = {Science},
  volume  = {373},
  number  = {6557},
  pages   = {871--876},
  year    = {2021},
  doi     = {10.1126/science.abj8754}
}

@article{vankempen2024foldseek,
  author  = {van Kempen, Michel and Kim, Stephanie S. and Tumescheit, Charlotte and Mirdita, Milot and Lee, Jeongjae and Gilchrist, Cameron L. M. and S{\"o}ding, Johannes and Steinegger, Martin},
  title   = {Fast and accurate protein structure search with {Foldseek}},
  journal = {Nature Biotechnology},
  volume  = {42},
  number  = {2},
  pages   = {243--246},
  year    = {2024},
  doi     = {10.1038/s41587-023-01773-0}
}

@inproceedings{su2024saprot,
  author    = {Su, Jin and Han, Chenchen and Zhou, Yuyang and Shan, Junjie and Zhou, Xibin and Yuan, Fajie},
  title     = {{SaProt}: Protein Language Modeling with Structure-aware Vocabulary},
  booktitle = {International Conference on Learning Representations (ICLR)},
  year      = {2024},
  url       = {https://openreview.net/forum?id=6MRm3G4NiU}
}

@article{heinzinger2024prostt5,
  author  = {Heinzinger, Michael and Weissenow, Konstantin and Sanchez, Joaquin Gomez and Henkel, Adrian and Mirdita, Milot and Steinegger, Martin and Rost, Burkhard},
  title   = {Bilingual language model for protein sequence and structure},
  journal = {NAR Genomics and Bioinformatics},
  volume  = {6},
  number  = {4},
  pages   = {lqae150},
  year    = {2024},
  doi     = {10.1093/nargab/lqae150}
}

@inproceedings{gao2024foldtoken,
  author    = {Gao, Zhangyang and Tan, Cheng and Wang, Jue and Huang, Yufei and Wu, Lirong and Li, Stan Z.},
  title     = {{FoldToken}: Learning Protein Language via Vector Quantization and Beyond},
  booktitle = {Proceedings of the AAAI Conference on Artificial Intelligence},
  volume    = {39},
  pages     = {219--227},
  year      = {2025},
  doi       = {10.1609/aaai.v39i1.31998}
}

@inproceedings{wang2024dplm,
  author    = {Wang, Xinyou and Zheng, Zaixiang and Ye, Fei and Xue, Dongyu and Huang, Shujian and Gu, Quanquan},
  title     = {Diffusion Language Models Are Versatile Protein Learners},
  booktitle = {Proceedings of the 41st International Conference on Machine Learning},
  series    = {Proceedings of Machine Learning Research},
  volume    = {235},
  pages     = {52309--52333},
  publisher = {PMLR},
  year      = {2024},
  url       = {https://proceedings.mlr.press/v235/wang24ct.html}
}

@inproceedings{vandenoord2017vqvae,
  author    = {van den Oord, Aaron and Vinyals, Oriol and Kavukcuoglu, Koray},
  title     = {Neural Discrete Representation Learning},
  booktitle = {Advances in Neural Information Processing Systems},
  volume    = {30},
  publisher = {Curran Associates, Inc.},
  year      = {2017},
  url       = {https://proceedings.neurips.cc/paper_files/paper/2017/hash/7a98af17e63a0ac09ce2e96d03992fbc-Abstract.html}
}

@article{kirkpatrick2017forgetting,
  author  = {Kirkpatrick, James and Pascanu, Razvan and Rabinowitz, Neil and others},
  title   = {Overcoming catastrophic forgetting in neural networks},
  journal = {Proceedings of the National Academy of Sciences},
  volume  = {114},
  number  = {13},
  pages   = {3521--3526},
  year    = {2017},
  doi     = {10.1073/pnas.1611835114}
}

@article{luo2023forgetting,
  author  = {Luo, Yun and Yang, Zhen and Meng, Fandong and Li, Yafu and Zhou, Jie and Zhang, Yue},
  title   = {An Empirical Study of Catastrophic Forgetting in Large Language Models During Continual Fine-Tuning},
  journal = {IEEE Transactions on Audio, Speech and Language Processing},
  volume  = {33},
  pages   = {3776--3786},
  year    = {2025},
  doi     = {10.1109/TASLPRO.2025.3606231}
}

@inproceedings{wang2024mmlupro,
  author    = {Wang, Yubo and Ma, Xueguang and Zhang, Ge and others},
  title     = {{MMLU-Pro}: A More Robust and Challenging Multi-Task Language Understanding Benchmark},
  booktitle = {Advances in Neural Information Processing Systems (Datasets and Benchmarks Track)},
  volume    = {37},
  pages     = {95266--95290},
  publisher = {Curran Associates, Inc.},
  year      = {2024}
}

@article{zhang2005tmalign,
  author  = {Zhang, Y. and Skolnick, J.},
  title   = {{TM-align}: a protein structure alignment algorithm based on the {TM-score}},
  journal = {Nucleic Acids Research},
  volume  = {33},
  number  = {7},
  pages   = {2302--2309},
  year    = {2005},
  doi     = {10.1093/nar/gki524}
}

@article{steinegger2017mmseqs2,
  author  = {Steinegger, Martin and S{\"o}ding, Johannes},
  title   = {{MMseqs2} enables sensitive protein sequence searching for the analysis of massive data sets},
  journal = {Nature Biotechnology},
  volume  = {35},
  number  = {11},
  pages   = {1026--1028},
  year    = {2017},
  doi     = {10.1038/nbt.3988}
}

@article{robin2021cameo,
  author  = {Robin, Xavier and Haas, Juergen and Gumienny, Rafal and Smolinski, Anna and Tauriello, Gerardo and Schwede, Torsten},
  title   = {{Continuous Automated Model EvaluatiOn (CAMEO)}---Perspectives on the future of fully automated evaluation of structure prediction methods},
  journal = {Proteins: Structure, Function, and Bioinformatics},
  volume  = {89},
  number  = {12},
  pages   = {1977--1986},
  year    = {2021},
  doi     = {10.1002/prot.26213}
}

@article{ma2025prottex,
  title   = {{ProtTeX}: Structure-In-Context Reasoning and Editing of Proteins with Large Language Models},
  author  = {Ma, Zicheng and Fan, Chuanliu and Wang, Zhicong and others},
  journal = {arXiv preprint arXiv:2503.08179},
  year    = {2025}
}

@inproceedings{yuan2025structtokenbench,
  title     = {Protein Structure Tokenization: Benchmarking and New Recipe},
  author    = {Yuan, Xinyu and Wang, Zichen and Collins, Marcus and Rangwala, Huzefa},
  booktitle = {Proceedings of the 42nd International Conference on Machine Learning},
  series    = {PMLR},
  volume    = {267},
  year      = {2025},
  note      = {arXiv:2503.00089}
}

@inproceedings{wei2022cot,
  title     = {Chain-of-Thought Prompting Elicits Reasoning in Large Language Models},
  author    = {Wei, Jason and Wang, Xuezhi and Schuurmans, Dale and Bosma, Maarten and Ichter, Brian and Xia, Fei and Chi, Ed H. and Le, Quoc V. and Zhou, Denny},
  booktitle = {Advances in Neural Information Processing Systems},
  year      = {2022}
}

@misc{esm3,
  title        = {ESM3 structure tokenizer},
  howpublished = {\url{https://github.com/evolutionaryscale/esm}},
  note         = {Accessed 2026; exact commit and weight checksums recorded in the released archive}
}

@misc{dplm,
  title        = {DPLM-2 structure tokenizer},
  howpublished = {\url{https://github.com/bytedance/dplm.git}, commit \texttt{8a2e15e534}},
  note         = {Accessed 2026; exact commit and weight checksums recorded in the released archive}
}

@misc{foldtoken4,
  title        = {FoldToken4},
  howpublished = {\url{https://github.com/A4Bio/FoldToken_open.git}, commit \texttt{fed13c4d11}},
  note         = {Accessed 2026; exact commit and weight checksums recorded in the released archive}
}

@misc{profit,
  title        = {ProFiT},
  howpublished = {\url{https://github.com/QDKStorm/ProFiT.git}, commit \texttt{be4ff7cb6b}},
  note         = {Accessed 2026; exact commit and weight checksums recorded in the released archive}
}

@misc{stb,
  title        = {StructTokenBench (AminoAseed, VanillaVQ)},
  howpublished = {\url{https://github.com/KatarinaYuan/StructTokenBench.git}, commit \texttt{2681b240f6}},
  note         = {Accessed 2026; exact commit and weight checksums recorded in the released archive}
}

@misc{gcpvq,
  title        = {GCP-VQVAE},
  howpublished = {\url{https://github.com/mahdip72/vq_encoder_decoder.git}, commit \texttt{68c4c284fe}},
  note         = {Accessed 2026; exact commit and weight checksums recorded in the released archive}
}

@article{gaujac2024language,
  title   = {Learning the Language of Protein Structure},
  author  = {Gaujac, Benoit and Don{\`a}, J{\'e}r{\'e}mie and Liu, Liviu and
             Sarkissian, Timothy and Beguir, Karim and Pierrot, Thomas},
  journal = {arXiv preprint arXiv:2405.15840},
  year    = {2024}
}

@article{zhang2024aidost,
  title   = {Balancing Locality and Reconstruction in Protein Structure Tokenizer},
  author  = {Zhang, Jiayou and Yang, Barthelemy Meynard-Piganeau and
             Gong, Jingyi and Fan, Xingyi and Ellington, Travis and
             Xing, Eric P. and Tan, Le Song},
  journal = {bioRxiv},
  year    = {2024},
  doi     = {10.1101/2024.12.02.626366}
}

@article{cao2025protokens,
  title   = {ProTokens: A Machine-Learned Language for Compact and Informative
             Encoding of Protein 3D Structures},
  author  = {Cao, Xiaohan and Feng, Zhicheng and Zhang, Zhengxin and Gao, Yi Qin},
  journal = {Chemical Science},
  year    = {2025},
  doi     = {10.1039/D5SC02055G}
}

@inproceedings{li2024prosst,
  title     = {{ProSST}: Protein Language Modeling with Quantized Structure and
               Disentangled Attention},
  author    = {Li, Mingchen and Tan, Yang and Ma, Xinzhu and Zhong, Bozitao and
               Yu, Huiqun and Zhou, Ziyi and Ouyang, Wanli and Zhou, Bingxin and
               Tan, Pan and Hong, Liang},
  booktitle = {Advances in Neural Information Processing Systems (NeurIPS)},
  volume    = {37},
  year      = {2024}
}

@article{hardy2026protok,
  title   = {{ProTok}: Unified Protein Tokenization via Modular Multi-Codebook
             Quantization},
  author  = {Hardy, Romain and Zhang, Xiaoman and Rajpurkar, Pranav},
  journal = {arXiv preprint arXiv:2610.18481},
  year    = {2026}
}

@inproceedings{yu2022vitvqgan,
  title     = {Vector-Quantized Image Modeling with Improved {VQGAN}},
  author    = {Yu, Jiahui and Li, Xin and Koh, Jing Yu and Zhang, Han and
               Pang, Ruoming and Qin, James and Ku, Alexander and Xu, Yuanzhong and
               Baldridge, Jason and Wu, Yonghui},
  booktitle = {International Conference on Learning Representations (ICLR)},
  year      = {2022}
}

@inproceedings{yu2024magvit2,
  title     = {Language Model Beats Diffusion: Tokenizer is Key to Visual Generation},
  author    = {Yu, Lijun and Lezama, Jos{\'e} and Gundavarapu, Nitesh B. and
               Versari, Luca and Sohn, Kihyuk and Minnen, David and Cheng, Yong and
               Gupta, Agrim and Gu, Xiuye and Hauptmann, Alexander G. and
               Gong, Boqing and Yang, Ming-Hsuan and Essa, Irfan and
               Ross, David A. and Jiang, Lu},
  booktitle = {International Conference on Learning Representations (ICLR)},
  year      = {2024}
}

@inproceedings{mentzer2024fsq,
  title     = {Finite Scalar Quantization: {VQ-VAE} Made Simple},
  author    = {Mentzer, Fabian and Minnen, David and Agustsson, Eirikur and
               Tschannen, Michael},
  booktitle = {International Conference on Learning Representations (ICLR)},
  year      = {2024}
}

@inproceedings{chang2022maskgit,
  title     = {{MaskGIT}: Masked Generative Image Transformer},
  author    = {Chang, Huiwen and Zhang, Han and Jiang, Lu and Liu, Ce and
               Freeman, William T.},
  booktitle = {IEEE/CVF Conference on Computer Vision and Pattern Recognition (CVPR)},
  year      = {2022}
}

@article{zeghidour2022soundstream,
  title   = {{SoundStream}: An End-to-End Neural Audio Codec},
  author  = {Zeghidour, Neil and Luebs, Alejandro and Omran, Ahmed and
             Skoglund, Jan and Tagliasacchi, Marco},
  journal = {IEEE/ACM Transactions on Audio, Speech, and Language Processing},
  volume  = {30},
  pages   = {495--507},
  year    = {2022}
}

@article{defossez2023encodec,
  title   = {High Fidelity Neural Audio Compression},
  author  = {D{\'e}fossez, Alexandre and Copet, Jade and Synnaeve, Gabriel and
             Adi, Yossi},
  journal = {Transactions on Machine Learning Research (TMLR)},
  year    = {2023}
}

@article{dietterich1995ecoc,
  title   = {Solving Multiclass Learning Problems via Error-Correcting Output Codes},
  author  = {Dietterich, Thomas G. and Bakiri, Ghulum},
  journal = {Journal of Artificial Intelligence Research},
  volume  = {2},
  pages   = {263--286},
  year    = {1995}
}

@article{hamming1950,
  title   = {Error Detecting and Error Correcting Codes},
  author  = {Hamming, Richard W.},
  journal = {The Bell System Technical Journal},
  volume  = {29},
  number  = {2},
  pages   = {147--160},
  year    = {1950}
}

@inproceedings{bengio2015scheduled,
  title     = {Scheduled Sampling for Sequence Prediction with Recurrent Neural Networks},
  author    = {Bengio, Samy and Vinyals, Oriol and Jaitly, Navdeep and Shazeer, Noam},
  booktitle = {Advances in Neural Information Processing Systems (NeurIPS)},
  year      = {2015}
}

@inproceedings{ranzato2016sequence,
  title     = {Sequence Level Training with Recurrent Neural Networks},
  author    = {Ranzato, Marc'Aurelio and Chopra, Sumit and Auli, Michael and
               Zaremba, Wojciech},
  booktitle = {International Conference on Learning Representations (ICLR)},
  year      = {2016}
}

@inproceedings{holtzman2020degeneration,
  title     = {The Curious Case of Neural Text Degeneration},
  author    = {Holtzman, Ari and Buys, Jan and Du, Li and Forbes, Maxwell and
               Choi, Yejin},
  booktitle = {International Conference on Learning Representations (ICLR)},
  year      = {2020}
}

@inproceedings{welleck2020unlikelihood,
  title     = {Neural Text Generation With Unlikelihood Training},
  author    = {Welleck, Sean and Kulikov, Ilia and Roller, Stephen and
               Dinan, Emily and Cho, Kyunghyun and Weston, Jason},
  booktitle = {International Conference on Learning Representations (ICLR)},
  year      = {2020}
}

@inproceedings{wang2023selfconsistency,
  title     = {Self-Consistency Improves Chain of Thought Reasoning in Language Models},
  author    = {Wang, Xuezhi and Wei, Jason and Schuurmans, Dale and Le, Quoc and
               Chi, Ed and Narang, Sharan and Chowdhery, Aakanksha and Zhou, Denny},
  booktitle = {International Conference on Learning Representations (ICLR)},
  year      = {2023}
}

@article{kabsch1983dssp,
  title   = {Dictionary of Protein Secondary Structure: Pattern Recognition of
             Hydrogen-Bonded and Geometrical Features},
  author  = {Kabsch, Wolfgang and Sander, Christian},
  journal = {Biopolymers},
  volume  = {22},
  number  = {12},
  pages   = {2577--2637},
  year    = {1983}
}

@article{zemla2003lga,
  title   = {{LGA}: A Method for Finding 3D Similarities in Protein Structures},
  author  = {Zemla, Adam},
  journal = {Nucleic Acids Research},
  volume  = {31},
  number  = {13},
  pages   = {3370--3374},
  year    = {2003}
}

@article{buttenschoen2024posebusters,
  title   = {{PoseBusters}: {AI}-Based Docking Methods Fail to Generate
             Physically Valid Poses or Generalise to Novel Sequences},
  author  = {Buttenschoen, Martin and Morris, Garrett M. and Deane, Charlotte M.},
  journal = {Chemical Science},
  volume  = {15},
  number  = {9},
  pages   = {3130--3139},
  year    = {2024}
}

@article{xu2025foldbench,
  title   = {Benchmarking All-Atom Biomolecular Structure Prediction with {FoldBench}},
  author  = {Xu, Sheng and Zhang, Odin and Wang, Yuchen and others},
  journal = {Nature Communications},
  year    = {2025},
  doi     = {10.1101/2025.05.22.655600}
}

@article{kryshtafovych2026casp16,
  title   = {Progress and Bottlenecks for Deep Learning in Computational Structure
             Biology: {CASP} Round {XVI}},
  author  = {Kryshtafovych, Andriy and Schwede, Torsten and Topf, Maya and
             Fidelis, Krzysztof and Moult, John},
  journal = {Proteins: Structure, Function, and Bioinformatics},
  volume  = {94},
  number  = {1},
  pages   = {5--14},
  year    = {2026}
}

@article{yuan2026casp16monomer,
  title   = {{CASP16} Protein Monomer Structure Prediction Assessment},
  author  = {Yuan, Qianmu and Tian, Chengxin and Zhang, Yang},
  journal = {Proteins: Structure, Function, and Bioinformatics},
  volume  = {94},
  number  = {1},
  pages   = {86--105},
  year    = {2026}
}

@inproceedings{rao2021contacts,
  title     = {Transformer Protein Language Models are Unsupervised Structure Learners},
  author    = {Rao, Roshan and Meier, Joshua and Sercu, Tom and Ovchinnikov, Sergey and
               Rives, Alexander},
  booktitle = {International Conference on Learning Representations (ICLR)},
  year      = {2021}
}

@article{efron1979bootstrap,
  title   = {Bootstrap Methods: Another Look at the Jackknife},
  author  = {Efron, Bradley},
  journal = {The Annals of Statistics},
  volume  = {7},
  number  = {1},
  pages   = {1--26},
  year    = {1979}
}

@article{mcnemar1947,
  title   = {Note on the Sampling Error of the Difference Between Correlated
             Proportions or Percentages},
  author  = {McNemar, Quinn},
  journal = {Psychometrika},
  volume  = {12},
  number  = {2},
  pages   = {153--157},
  year    = {1947}
}

@inproceedings{henderson2018deeprl,
  title     = {Deep Reinforcement Learning That Matters},
  author    = {Henderson, Peter and Islam, Riashat and Bachman, Philip and
               Pineau, Joelle and Precup, Doina and Meger, David},
  booktitle = {AAAI Conference on Artificial Intelligence},
  year      = {2018}
}

@inproceedings{dodge2019showyourwork,
  title     = {Show Your Work: Improved Reporting of Experimental Results},
  author    = {Dodge, Jesse and Gururangan, Suchin and Card, Dallas and
               Schwartz, Roy and Smith, Noah A.},
  booktitle = {Conference on Empirical Methods in Natural Language Processing (EMNLP)},
  year      = {2019}
}

@inproceedings{reimers2017seeds,
  title     = {Reporting Score Distributions Makes a Difference: Performance Study of
               {LSTM}-networks for Sequence Tagging},
  author    = {Reimers, Nils and Gurevych, Iryna},
  booktitle = {Conference on Empirical Methods in Natural Language Processing (EMNLP)},
  year      = {2017}
}

@inproceedings{gururangan2020dapt,
  title     = {Don't Stop Pretraining: Adapt Language Models to Domains and Tasks},
  author    = {Gururangan, Suchin and Marasovi{\'c}, Ana and Swayamdipta, Swabha and
               Lo, Kyle and Beltagy, Iz and Downey, Doug and Smith, Noah A.},
  booktitle = {Annual Meeting of the Association for Computational Linguistics (ACL)},
  year      = {2020}
}

@article{kaplan2020scaling,
  title   = {Scaling Laws for Neural Language Models},
  author  = {Kaplan, Jared and McCandlish, Sam and Henighan, Tom and Brown, Tom B. and
             Chess, Benjamin and Child, Rewon and Gray, Scott and Radford, Alec and
             Wu, Jeffrey and Amodei, Dario},
  journal = {arXiv preprint arXiv:2001.08361},
  year    = {2020}
}

@inproceedings{hoffmann2022chinchilla,
  title     = {Training Compute-Optimal Large Language Models},
  author    = {Hoffmann, Jordan and Borgeaud, Sebastian and Mensch, Arthur and others},
  booktitle = {Advances in Neural Information Processing Systems (NeurIPS)},
  year      = {2022}
}

\clearpage

\appendix
{\titlefont\sffamily\seedblue{Appendix}\par}
\vskip 4mm

\section{Measurement incidents and what they cost}
\label{app:incidents}

Every control in Section~\ref{sec:method} exists because its absence produced a wrong number
first. We record the four incidents because the resulting rules are the transferable part of
this work, and because each of them was caught late enough to be worth describing.

\paragraph{Reporting a capability score that was below the floor.}
In an earlier round we reported a greedy TM of 0.168 as a capability result. The random-token
floor for that tokenizer, measured afterwards on the same targets at the same sample size, was
0.185. The model's output was worse than random input through the same decoder. Two facts
needed to detect this were already written in that report: that local geometry was guaranteed
by the decoder rather than by the model, and that greedy decoding repeated the same token at
about 25\% of positions. Neither had been followed to its conclusion.
\emph{Rule adopted:} a floor is part of the result, not a follow-up, and an observed anomaly is
pursued to a conclusion rather than noted.

\paragraph{Scoring the gift.}
The prefix-rescue script's header stated that it scored only the unforced suffix; its
implementation called the general scorer, which scores the whole chain. Because the forced
prefix is copied ground truth, the reported effect was inflated 2.7-fold at $q=0.25$ and more
at larger $q$. The script's own log line said "whole-chain TM"; the header and the log were
never read against each other. \emph{Rule adopted:} when a comment and a log disagree, the log
is the specification; suffix-only scoring now runs automatically at the end of that script.

\paragraph{Numbers with no file behind them.}
The error-geometry measurements underlying Section~\ref{sec:mechanism} initially existed only
as terminal output. They were recomputed and archived as JSON, at which point extending them
from one checkpoint to all four cost nothing, and produced the saturation result that
explains the null interventions. \emph{Rule adopted:} every number in a report resolves to an
archived file, and analysis scripts write their results rather than printing them.

\paragraph{An intervention too gentle to test anything.}
The first bit-supervision variant used $\epsilon = 0.2$, chosen to match an existing
label-smoothing convention. It moved the error Hamming ratio from 0.506 to 0.507. A
pre-registered mechanism check caught this and voided the run; without it, a clean-looking
null would have been published as evidence that bit supervision does not help.
\emph{Rule adopted:} a conservative hyperparameter is not a safe one, and every intervention
ships with a pre-registered check that it did something.

\paragraph{Self-corrections.}
Four conclusions were revised during the study as data accumulated: the role of codebook size
(revised four times before settling on "no effect that survives the noise floor"), the relative
standing of two tokenizers (three times), a diagnosis of an encoding bottleneck that was wrong
about which resource was saturated, and a claimed result that held on only one of three
evaluation sets and was withdrawn.
\emph{Rule adopted:} the report records when each judgement was made and on what data.

\section{Reproducibility}
\label{app:repro}

\paragraph{What is released.}
All 131 evaluation cells are archived, each containing the summary metrics, per-target scores,
the reference-free panel, and the \emph{raw generated token streams}. The last item is what
makes the archive useful: the geometry panel of Section~\ref{sec:results}, the error-geometry
trajectory of Section~\ref{sec:mechanism} and the suffix-only rescue scores of
Section~\ref{sec:negative} were all computed after the fact from archived generations, on
checkpoints trained weeks earlier, without loading a single model. Also archived: 11 training
curves at per-step granularity, the cross-tokenizer scorecard containing every floor, ceiling,
stability and tolerance measurement, the data-construction statistics for each arm's mixture,
and the per-item MMLU-Pro predictions.

\paragraph{Third-party components.}
Each tokenizer is pinned to a repository commit and its weights to a SHA-256 checksum, both
recorded in the archive's provenance directory. Model checkpoints
(four exports per arm, roughly 17\,GB each) are not distributed; the training mixtures can be
rebuilt from the recorded construction parameters, and the checkpoints cannot, which is noted
in the archive.

\paragraph{Known gaps.}
ESM3's four compute points in the released grid come from a rerun under this paper's pipeline
rather than from the original run, which had no 1.5B export and whose training checkpoints had
been rotated away; the original run's three points are kept separately and used only for the
noise estimate. VanillaVQ's 1.5B and 2B points come from resuming its 1B checkpoint rather
than from a continuous run. Neither affects the comparisons made in this paper, and both are
marked in the archive.

\section{Full result tables}
\label{app:tables}

\begin{table}[h]
  \centering
  \caption{The complete grid: greedy TM-score for every tokenizer, compute budget and
  evaluation set (96 cells). Within-column comparisons smaller than 0.013 TM are below the
  measured run-to-run noise floor (Section~\ref{sec:method}).}
  \label{tab:grid}
  \small
  \begin{tabular}{llrrr}
    \toprule
    Tokenizer & Budget & FoldBench334 $\uparrow$ & CASP16 $\uparrow$ & CAMEO2025 $\uparrow$ \\
    \midrule
    DPLM-2 & 0.5B & 0.5919 & 0.4888 & 0.6791 \\
     & 1B & 0.6455 & 0.5289 & 0.7538 \\
     & 1.5B & 0.6564 & 0.5521 & 0.7787 \\
     & 2B & 0.6623 & 0.5689 & 0.7999 \\
    \midrule
    ESM3 & 0.5B & 0.5478 & 0.4577 & 0.6562 \\
     & 1B & 0.6025 & 0.4910 & 0.7455 \\
     & 1.5B & 0.6155 & 0.5123 & 0.7836 \\
     & 2B & 0.6162 & 0.5482 & 0.7946 \\
    \midrule
    GCP-VQVAE & 0.5B & 0.3847 & 0.3324 & 0.4540 \\
     & 1B & 0.4771 & 0.4102 & 0.6102 \\
     & 1.5B & 0.5179 & 0.4263 & 0.6810 \\
     & 2B & 0.5254 & 0.4407 & 0.7106 \\
    \midrule
    FoldToken4-l12 & 0.5B & 0.4101 & 0.3617 & 0.4449 \\
     & 1B & 0.4841 & 0.4299 & 0.5827 \\
     & 1.5B & 0.5283 & 0.4604 & 0.6487 \\
     & 2B & 0.5281 & 0.4586 & 0.6761 \\
    \midrule
    FoldToken4-l10 & 0.5B & 0.3740 & 0.3556 & 0.4185 \\
     & 1B & 0.4639 & 0.4019 & 0.5395 \\
     & 1.5B & 0.5129 & 0.4521 & 0.6344 \\
     & 2B & 0.5270 & 0.4532 & 0.6645 \\
    \midrule
    ProFiT & 0.5B & 0.3303 & 0.3033 & 0.3611 \\
     & 1B & 0.4020 & 0.3728 & 0.4923 \\
     & 1.5B & 0.4325 & 0.3885 & 0.5446 \\
     & 2B & 0.4467 & 0.3845 & 0.5678 \\
    \midrule
    AminoAseed & 0.5B & 0.3692 & 0.2807 & 0.4170 \\
     & 1B & 0.3880 & 0.3094 & 0.4581 \\
     & 1.5B & 0.3941 & 0.2996 & 0.4668 \\
     & 2B & 0.3952 & 0.3036 & 0.4727 \\
    \midrule
    VanillaVQ & 0.5B & 0.3672 & 0.3291 & 0.3767 \\
     & 1B & 0.3808 & 0.3252 & 0.3987 \\
     & 1.5B & 0.3769 & 0.3246 & 0.3992 \\
     & 2B & 0.3781 & 0.3272 & 0.4008 \\
    \bottomrule
  \end{tabular}
\end{table}

\begin{table}[h]
  \centering
  \caption{Tokenizer scorecard, measured on FoldBench334 without training any model.
  $S_1$ is the reconstruction ceiling, $S_4$ the uniform-random floor, both through that
  tokenizer's own decoder. Bond-validity columns are the fraction of consecutive CA--CA
  distances inside $[3.6,4.0]$\,\AA{} at the ceiling and at the floor. $S_3$ is the TM retained
  after replacing 10\% of ground-truth tokens at random. $S_5$ is prefix consistency.}
  \label{tab:scorecard}
  \small
  \begin{tabular}{lrrrrrrr}
    \toprule
    & & \multicolumn{2}{c}{TM} & \multicolumn{2}{c}{CA--CA validity} & & \\
    \cmidrule(lr){3-4}\cmidrule(lr){5-6}
    Tokenizer & $K$ & $S_1$ $\uparrow$ & $S_4$ $\downarrow$ & ceiling & floor & $S_3$@10\% $\uparrow$ & $S_5$ $\uparrow$ \\
    \midrule
    DPLM-2 & 8192 & 0.9519 & 0.1766 & 0.996 & 0.806 & 0.890 & 0.2100 \\
    ESM3 & 4096 & 0.9886 & 0.1847 & 0.975 & 0.087 & 0.671 & 0.5927 \\
    GCP-VQVAE & 4096 & 0.9891 & 0.1029 & 0.991 & 0.149 & 0.839 & 0.0490 \\
    FoldToken4-l12 & 4096 & 0.9620 & 0.1436 & 0.903 & 0.026 & 0.853 & 0.0205 \\
    FoldToken4-l10 & 1024 & 0.9594 & 0.1508 & 0.891 & 0.028 & 0.849 & 0.0284 \\
    ProFiT & 8192 & 0.9902 & 0.1680 & 0.995 & 0.793 & 0.912 & 0.0011 \\
    AminoAseed & 512 & 0.4816 & 0.1847 & 0.337 & 0.126 & 0.445 & 0.5375 \\
    VanillaVQ & 512 & 0.4226 & 0.1885 & 0.246 & 0.083 & 0.407 & 0.6545 \\
    \bottomrule
  \end{tabular}
\end{table}

\begin{table}[h]
  \centering
  \caption{General-capability guardrail: MMLU-Pro on a pre-registered 350-item stratified
  subset, official 5-shot chain-of-thought protocol, at the 2B-token checkpoint.
  No pairwise difference is significant (exact McNemar); see Section~\ref{sec:guardrail}.}
  \label{tab:mmlu}
  \small
  \begin{tabular}{lrrrrr}
    \toprule
    Arm & TM @2B & Accuracy $\uparrow$ & $\Delta$ vs base & Parse failures $\downarrow$ & Mean CoT tokens \\
    \midrule
    base model & -- & 0.7743 & -- & 0.29\% & 661 \\
    DPLM-2 & 0.6623 & 0.7571 & -1.71pp & 0.86\% & 371 \\
    ESM3 & 0.6162 & 0.7629 & -1.14pp & 0.29\% & 360 \\
    GCP-VQVAE & 0.5254 & 0.7457 & -2.86pp & 0.29\% & 379 \\
    FoldToken4-l12 & 0.5281 & 0.7371 & -3.71pp & 0.57\% & 373 \\
    FoldToken4-l10 & 0.5270 & 0.7371 & -3.71pp & 0.00\% & 359 \\
    ProFiT & 0.4467 & 0.7257 & -4.86pp & 0.00\% & 370 \\
    AminoAseed & 0.3952 & 0.7371 & -3.71pp & 0.29\% & 331 \\
    \bottomrule
  \end{tabular}
\end{table}

\end{document}